\documentclass[prb,twocolumn,color,superscriptaddress,showpacs]{revtex4}
\usepackage{graphicx}
\usepackage{color,amsmath,graphicx,latexsym}
\usepackage{epsfig}
\definecolor{li}{rgb}{0,0,1}
\begin{document}
\title{ 
Temperature dependent optical conductivity of undoped cuprates
with weak exchange}
\author{J.~M\'{a}lek}
\affiliation{Leibniz-Institut f\"ur Festk\"orper- und 
Werkstoffforschung
Dresden, P.O.\ Box 270116, D-01171 Dresden, Germany}
\affiliation{Institute of Physics,  
ASCR, Na Slovance 2, CZ-18221 Praha 8, 
Czech Republic}
\author{S.-L.\ Drechsler$^{*}$}
\affiliation{Leibniz-Institut f\"ur Festk\"orper- und 
Werkstoffforschung
Dresden, P.O.\ Box 270116, D-01171 Dresden, Germany}

\author{U.\ Nitzsche}
\affiliation{Leibniz-Institut f\"ur Festk\"orper- und 
Werkstoffforschung
Dresden, P.O.\ Box 270116, D-01171 Dresden, Germany}

\author{H.\ Rosner}
\affiliation{Max-Planck-Institut f\"ur Chemische Physik fester Stoffe,
Dresden, Germany}
\author{H.\ Eschrig}
\affiliation{Leibniz-Institut f\"ur Festk\"orper- und 
Werkstoffforschung
Dresden, P.O.\ Box 270116, D-01171 Dresden, Germany}

\date{\today}
\begin{abstract}
\noindent 
\textcolor{black}{The optical conductivity $\sigma(\omega)$ is calculated
at finite temperature $T$ for CuO$_2$ chain clusters
within  a $pd$-Hubbard model.
Data }at $T$~=~300~K for
Li$_2$CuO$_2$ are reanalyzed within  this approach. 
The relative weights of Zhang-Rice  singlet and triplet 
charge excitations near 2.5 and 4~eV, respectively, depend 
strongly on $T$, and   a 
rather dramatic dependence of $\sigma(\omega)$ on the
ratio of the 1$^{st}$ to 2$^{nd}$ neighbor exchange integrals
is predicted. 
On the basis of these results, information about
exchange interactions
\textcolor{black}{for frustrated edge-shared cuprates} 
can be obtained from 
\textcolor{black}{$T$-dependent} 
optical spectra.  \textcolor{black}{Our results are}
also relevant for   
magnetically weakly
coupled wide-gap insulators in general.
\end{abstract}
\maketitle

Standard wisdom on wide gap insulators says that their optical spectra
above 1~eV excitation energy are hardly affected by temperature $T$,
magnetic fields $H$, and by the magnetic nature of their ground state
(GS).  Moreover, spin and charge degrees of freedom are often
decoupled in one dimension (1D).  Here, we present exact theoretical
results qualitatively valid for several cuprates to be specified below
which prove just the opposite in all these respects.  Our study bases
on the fact that different magnetic states with different symmetries
obey different selection rules, which - in case of soft magnetic
materials~- can lead to sizable $T$-dependence of the optical
spectra.  For systems with small enough exchange integrals, the energy
difference between the GS and the excited magnetic states of the
system will be small enough so that they can be thermally
populated. This will cause a strong $T$-dependence of 
various response functions, here presented for the case of
the optical conductivity $\sigma(\omega)$.
For an illustration of our approach we consider Li$_2$CuO$_2$ 
being of current interest
\cite{mizuno,atzkern,drechsler07a,Vernay07,Learmonth07,sapina,neudert,hu,Graaf02,Hasan03,Kim04,Chung03},
structurally simple, and most importantly, where single crystal data
are available in a broad $\omega$-range \cite{mizuno,atzkern}.  This
system stands for a class of
frustrated spin-1/2 chain materials with
\textcolor{black}{both} small ferromagnetic (FM) NN Cu-Cu exchange
coupling $J_1$ and AFM next-nearest neighbor (NNN) Cu-Cu in-chain
exchange $J_2$ \cite{mizuno,drechsler07a} in terms of a 1D spin-1/2
Heisenberg model
\begin{equation}
{\cal{H}}_{\mbox{\tiny S}}=\sum_{i} J_1{\mathbf S}_i{\mathbf S}_{i+1}+
J_2{\mathbf S}_i{\mathbf S}_{i+2}+
J_3{\mathbf S}_i{\mathbf S}_{i+3}+... 
\end{equation}
In  Li$_2$CuO$_2$  
chains running
along the $b$-axis are
formed by the edge-sharing of CuO$_4$ plaquettes.
The most exciting puzzle  addressed here is  
the missing of the Zhang-Rice singlet peak (ZRS) 
in reflectivity and electron energy loss spectroscopy (EELS)
data at 
$T=$~300~K
\cite{mizuno,atzkern}.
Its detection in resonant inelastic x-ray  scattering (RIXS)
spectra is under debate, too \cite{Vernay07,Learmonth07}. There is also 
no consensus on
the role of two scenarios for the FM in-chain order 
below the N\'eel temperature $T_N=$~9~K. 
Scenario I is given by a dominant FM $J_1$
\cite{Graaf02} defined by 
the inequality
$\alpha=-J_2/J_1<\alpha_c\approx$~0.25, if $J_3$ is
weak \cite{drechsler07b}, where $\alpha_c$ marks the transition
from a high-spin FM to a low-spin
spiral-like ground state (GS). While in scenario II
it stems from a specific AFM inter-chain coupling
and a single chain would be at $\alpha$~$>$~$\alpha_c$  
\cite{mizuno,Vernay07,Whangbo07}.

A first microscopic theoretical description \cite{mizuno} of the magnetic 
susceptibility $\chi(T)$ and of   
$\sigma(\omega)$ at {\it zero temperature} on the basis of a 
Cu 3$d_{xy}$ O 2$p_{x,y}$ 
Hubbard model 
($x$ and $y$ along the $b$-  and $c$-axes, respectively, see Figs.~1,2 of 
Ref.~\onlinecite{mizuno})
predicts charge transfer excitation of a Cu hole into a
ZRS  near $\omega_{\mbox{\tiny ZRS}}$=~2.25~eV. Its
missing observation (at $T$=~300~K) 
was ascribed to poor resolution \cite{mizuno}, 
or to uncoupled CuO$_4$ units.\cite{atzkern}
We show, that a ZRS visible at low $T$
would be strongly suppressed at 300~K.
However, if the Hamiltonian parameters are chosen
to improve the description of 
O~1$s$ x-ray absorption (XAS) \cite{neudert},
optical, EELS \cite{mizuno,atzkern}, and RIXS data 
\cite{Learmonth07},
 then the ZRS absent at $T$~=~0,
appears with rising $T$ but is accompanied by a Zhang-Rice triplet 
(ZRT) contribution 
at $\omega_{\mbox{\tiny ZRT}}\approx$~
4~eV in $\sigma(\omega)$
(superimposed with further transitions). 
The $\sigma(\omega)$ data are well described this way.

The used Hubbard Hamiltonian reads
 (cf.\ also Ref.\ \onlinecite{mizuno})
\begin{equation}
\cal{H}_{\mbox{\tiny H}}=\cal{H}_{\mbox{\tiny k}}+\cal{H}_{\mbox{\tiny C}}+\cal{H}_{\mbox{\tiny ex}}.
\end{equation}
Its kinetic, Coulomb,  and exchange  contributions are
\begin{eqnarray}
\cal{H}_{\mbox{\tiny k}}&=&\sum_{i}\varepsilon_in_{i}
+\sum_{i,j,s}t_{ij}c^\dag_{i,s}c_{j,s}, 
\ n_{is}=c^\dag_{is}c_{is},\\
\cal{H}_{\mbox{\tiny C}}&=&\sum_{i}U_{i}n_{i\uparrow}n_{i\downarrow}
+\frac{1}{2}\sum_{i\neq j}V_{ij}n_{i}n_j, \ n_i=\sum_sn_{is},\\
\cal{H}_{\mbox{\tiny ex}}&=&\frac{1}{2}\sum_{i\neq j,ss'}K_{ij}\left(c^\dag_{is}
c^\dag_{js'}
c_{is'}c_{js}+ c^\dag_{is}c^\dag_{is'}
c_{js'}c_{js}\right),\\
\nonumber
\end{eqnarray}
where $i$ and $j$ run over all Cu-3$d_{xy}$ and O-2$p_{x,y}$ orbitals and
$s$ is the spin index.
Except for the $\varepsilon_i$ the Hamiltonian
parameters are the same as in Ref.\ \onlinecite{mizuno}:
$U_d =$8.5~eV, $U_{p}=$~4.1~eV for the intra-orbital 
and $V_{p_x,p_y}=U_{p}-2K_{p_x,p_y}=2.9$ eV
for the O onsite inter-orbital repulsion,
where $K_{p_x,p_y}=0.6$ eV
is the FM Hund's rule coupling.
As in Ref.\ 1 isotropy of the FM Cu-O exchange integrals
$K_{pd}=K_{p_xd}=K_{p_yd}=0.05$ eV 
was used for the sake of simplicity.
Polarized O 1$s$ XAS measurements 
\cite{neudert} with the electric field 
vector in $x$ and $y$ direction, respectively,   
indicated a nearly isotropic O $2p$-hole distribution in the $xy$-plane.
This indicates a condition 
$n_{p_x}\approx n_{p_y}$ within XAS error bars
for the O 2$p$ hole occupation numbers in  the GS. 
In order to achieve this despite the anisotropic CuO$_4$ 
plaquette geometry, $\varepsilon_{p_x}$-$\varepsilon_{p_y}=$~0.2~eV 
was taken as distinct from Ref.~1
where $\varepsilon_{p_x}$~=~$\varepsilon_{p_y}$
was assumed. In order to reproduce the first strong peak in
$\sigma(\omega)$ near 4.4$\pm$~0.2~eV (Figs.~1,2), the mean O 
onsite energy
$\Delta_{pd}=(\varepsilon_{p_x}+\varepsilon_{p_y})/2-\varepsilon_d$ 
has additionally been 
up-shifted by 0.5 to 3.7~eV.
To demonstrate the strikingly distinct $\sigma(\omega)$
caused by these \textcolor{black}{moderate}  
changes, calculations were performed with the 
$\varepsilon_i$-values of Ref.\ \onlinecite{mizuno}, too.
\textcolor{black}{Hereafter} these two models (2) are called
M1 \cite{mizuno} and M2 (above choice). We will show
that they lead to different magnetic GS yielding 
{\it different} $T$-behavior of $\sigma(\omega)$. 

Naturally, the dominating exchange integrals $J_1$ and $J_2$
 determine the spectrum of low-energy excited states $|\nu\rangle$ 
(spin excitations) of the spin model (1). We found the $J$-values  from 
projecting the Hamiltonian (2) onto (1).
At a given $T$ and possibly in the presence of a magnetic 
field $H$, the optical conductivity
$\sigma(\omega,H,T)$ of (2) 
is obtained from the $\sigma_\nu(\omega)$ with the initial 
\textcolor{black}{spin}
states $\mid \nu\rangle$ of (2) \cite{jaclic}:  
\begin{eqnarray}
\sigma(\omega,T)&=&\sum_\nu w_\nu (T)
\left[1-\exp(-\hbar\omega/k_{\mbox{\tiny B}}T)\right]
\sigma_\nu (\omega),\\
w_\nu &=&\frac{g_\nu \exp(-E_\nu/k_{\mbox{\tiny B}}T)}{\sum_{\nu'} g_{\nu'} \exp(-E_{\nu'}/k_{\mbox{\tiny B}}T)},\\
\nonumber
\end{eqnarray}
where for $H=0$  the spin degeneracy is $g_\nu=2S_\nu+1$.
For instance, for the largest cluster, Cu$_6$O$_{14}$, which
can still be handled by the exact diagonalization of the Hamiltonian (2)
there are 2$^6$=64 low-energy spin states: 
$5\times (S=0)$, $9\times (S=1)$, $5\times (S=2)$, and 
$1\times (S=3)$ multiplets.
Due to the large optical transition energies
$\hbar \omega \gg k_{\mbox{\tiny B}}T$
 the thermal occupation of the final states in
 (6) can be ignored.
 In case of an applied external magnetic field $H$ the $S\neq 0$
states with finite $S_z$ are Zeeman split which affects 
the Boltzmann  
probability $w_\nu=w_\nu (T,H)$ to find a cluster in a given spin state 
$(S_\nu,S_{z,\nu})$. In this case the $g_{\nu}$ are 
replaced by
\begin{eqnarray}
g_{\nu}(H,T)&=&1 +2\sum^{\nu_{max}}_{\nu'=1}\cosh\left[
\frac{\nu'g_{\mbox{\tiny L}}\mu_{\mbox{\tiny B}}H}{k_{\mbox{\tiny B}}T}\right],\\
g_\nu (H,T)&=&2\sum^{\nu_{max}}_{\nu'=1}\cosh\left[\frac{(2\nu'-1)
g_{\mbox{\tiny L}}\mu_{\mbox{\tiny B}}H}{2k_{\mbox{\tiny B}}T}\right],\\
\nonumber
\end{eqnarray}
for even and odd chains, respectively, where $g_{\mbox{\tiny L}}$ 
denotes the Land\'{e}-factor.
All response functions reported below were calculated 
using exact diagonalizations and 
the common continued fraction method for Cu$_n$O$_{2n+2}$ clusters.
The $\delta$-functions of the 
calculated $\sigma (\omega)$-spectra are convoluted with a Lorentzian
broadening of $\gamma_L$ = 0.35~eV at half width to compare them
with optical data.

Generally, the calculated $\sigma(\omega)$ for Cu$_n$O$_{2n+2}$
chain clusters at $H=T=0$  
exhibit a multiple peak structure: 
two well-pronounced peaks near 4 and 5 to 5.5~eV as 
shown in Fig.~1. Remarkably, there
are marked differences in $\sigma(\omega)$ 
between 2 and 4~eV depending whether the GS is low-spin
(i.e.\ a spin spiral) or high spin (i.e.\ FM).
For 
Li$_2$CuO$_2$, due to 
its
closeness 
to the critical 
point $\alpha_c$ (i.e.\ the FM-spiral transition), 
the issue 
depends sensitively on the  $J$-values governed by the details of
the parameters $\varepsilon_i$.
A main issue of our analysis is that while
the $\varepsilon_i$ values of M1 lead to a low-spin GS
(spiral, $\alpha > \alpha_c$) our choice M2 results in a high-spin GS
(FM, $\alpha < \alpha_c$). $\sigma(\omega)$
for a Cu$_6$O$_{14}$ chain with assumption of a 
high-spin GS (S=3) or a low-spin
GS (S=0,1) are shown on Fig.~2.
The height of the feature between 2 and 3~eV
depends on the cluster
size and on the total spin of the GS. Already a single CuO$_4$ unit shows
a double peak structure derived from 
transitions between an even parity  
singly occupied hybridized Cu 3$d$ O2$p$ state and
odd parity non-bonding O states. 
Like the GS, all states are spin doublets.
Since these transition energies do almost
coincide with 
the mentioned above "high-energy"
peaks for larger clusters, we regard them as
intra-plaquette excitations.  
Inter-plaquette transitions 
may occur starting from dimers $n$~=~2. If their GS
is an $S=0$ state or a low-spin state for larger clusters
(e.g.\ a doublet for odd chains), a peak appears between 2 and 3~eV.  
This transition is forbidden, if the GS 
is the maximal spin state: i.e.\
an $S=n/2$ FM state.
Since in the optical transition $S$ and $S_z$
are conserved, the final state for an excited
 dimer in the former
 case is again a 
singlet (low-spin state) with two holes at one of the two CuO$_4$ units,
one sitting mainly on Cu and one sitting mainly on O
(similar to a ZRS state on a CuO$_4^{-5}$ unit). 
This
transition is usually denoted as a 
ZRS excitation. In the  
FM case 
the excited state may contain
a triplet (ZRT) of two holes on one plaquette, which in the optical
excitation process 
occurs slightly below the main peak near 4 eV \cite{remarkenergy}.
The intensities of the ZRS and ZRT transitions exhibit some
positive finite size effect $\propto 1/n$ 
reflecting mainly the number of available
plaquettes for an inter-plaquette transition.
For a dimer with $J_1<0$ the GS is always a triplet
since $J_2$ is involved for $n\geq 3$, only.  
A detailed finite-size analysis will be given elsewhere.
The ZRT-energy $\approx$ 4~eV found above is in excellent 
agreement with 4.1~eV reported in a recent RIXS study
\cite{Learmonth07} but in sharp contrast with 
the 
assignment given in Ref.\ \onlinecite{Vernay07}
where the RIXS peak near 2.1~eV has been ascribed to 
the ZRS using a low-spin GS. 
Probably, the 2.1~eV-peak should be ascribed 
to $d$-$d$ excitations \cite{Kim04,Hasan03}. A 
similar feature has been found in CuGeO$_3$ (in $\sigma(\omega)$ 
at lower $\omega$~$\approx$~1.7~$\pm$ 0.1~eV) while the 
ZRS is clearly observed at 3.4 $\pm$ 0.2~eV 
\cite{Bondino07,Pagliara02}.
\begin{figure}[t]
\includegraphics[width=8.5cm,angle=0]{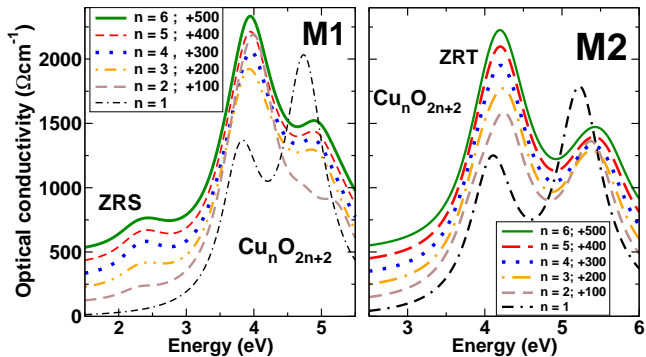}
\caption{(Color). \textcolor{black}{Finite size effect
for $\sigma(\omega)$. Left (right): GS 
given by the  lowest (highest) spin state.
The curves for various 
clusters are shifted
by (n-1)$\times$100$(\Omega cm)^{-1}$.
} 
}
\label{losstheo}
\end{figure}
\begin{figure}[b]
\includegraphics[width=7.2cm,angle=0]{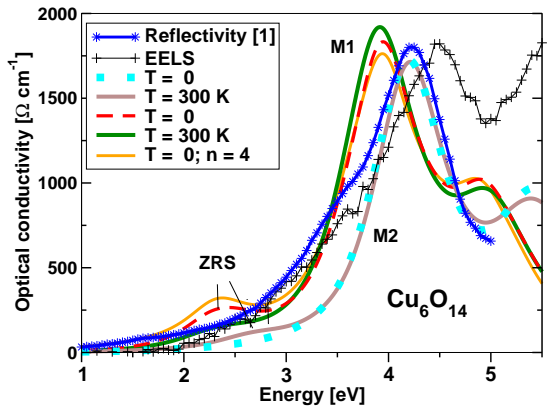}
\caption{(Color). \textcolor{black}{Experimental and 
theoretical optical in-chain conductivities of (2)
for $T=$0 and $T=300$ K using a low-spin($S$~=~0,1) GS (M1) and 
a high-spin ($S$~=~3) GS (M2).}}
\label{mizunocomp}
\end{figure}
\begin{figure}[t]
\includegraphics[width=8.2cm,angle=0]{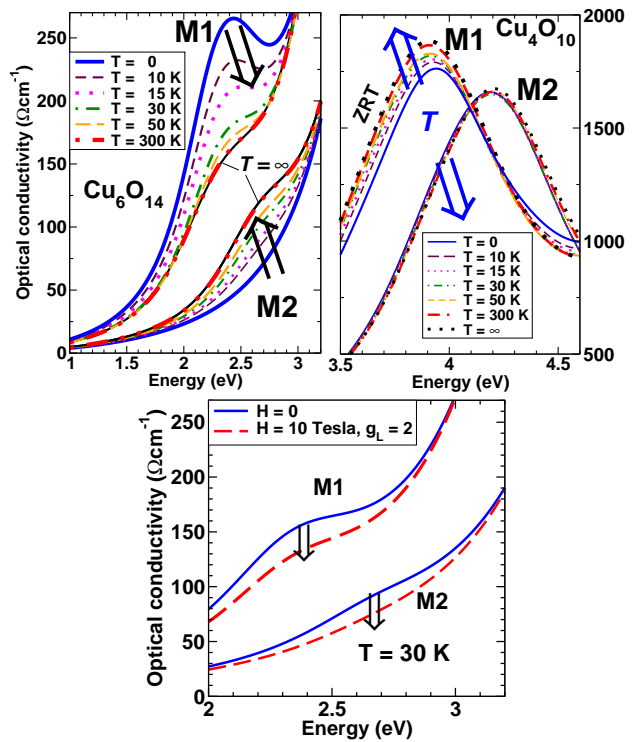}
\caption{\textcolor{black}{(Color). $T$-dependent
 optical conductivity at $H$~=~0} (upper)
and at $H\neq$~0 (lower) within the 
regions of ZRS
and ZRT transitions
for clusters
within a low-spin (M1) and a high-spin GS 
(M2). 
'$T=\infty$' means $ k_{\mbox{\tiny B}}T \gg \mid J_i \mid $, but 
still $\hbar \omega \geq 10^4 $~K$ \gg k_{\mbox{\tiny B}}T$.}
\label{tetratemp}
\end{figure}

Exact diagonalization of (2) yields excitation
energies $\omega_{\mbox{\tiny ZRS}}$~=~2.7~eV, 
$\omega_{\mbox{\tiny ZRT}}$~=~4~eV for M2
and 
$\omega_{\mbox{\tiny ZRS}}$~=~2.25~eV, $\omega_{\mbox{\tiny ZRT}}$~=~3.7~eV 
for M1.
For both models this is 
consistent with $\Delta_{\mbox{\tiny ZR}}$~$\approx$~1.3~eV
 obtained
as the distance between the lowest $S=0$ and the $S=1$ levels 
in a single CuO$_4^{-5}$ plaquette with two holes.
The $\sigma(\omega,T)$
obtained from (7) for both models are shown on Fig.~3, 
upper two panels. At 
300~K M1 and M2 yield the same qualitative behavior: the ZRS is largely 
suppressed. The main peak position of the experiment 
at ${\small \gtrsim}$~4eV is, however, much better reproduced by M2. We 
studied also the field dependence of $\sigma(\omega,T,H)$
(Fig.~3, lower panel) and found it similar in both models.
To repeat it, M1 results in $-J_2/J_1$~=~$\alpha$~$>$~$\alpha_c$
when projected onto the spin Hamiltonian (1) and hence 
in  a low-spin GS
while M2 resulting in $\alpha < \alpha_c$ 
and hence in a FM GS.
The reason for the reduction of $\alpha$
is mainly a reduction in superexchange due 
to the enhanced $\Delta_{pd}$-value.
As is clearly seen from Fig.~3, whether $\alpha<\alpha_c$
or $\alpha >\alpha_c$ can directly experimentally
be decided by low-$T$ studies of $\sigma(\omega)$ 
and such measurements
are strongly encouraged.
In this context we stress once more 
that our main result is
{\it not} simply the determination of a new, improved,
parameter 
set M2 in a literal sense,
 but the demonstration of an unique correlation
between the magnetic nature of the GS and the 
$T$-dependence of $\sigma(\omega)$. A refinement of M2 
when low-$T$ and $H\neq$~0 data 
will be available, a 
generalization of the Hamiltonian (2) 
to include further orbitals or interactions, its application to 
other cuprates
like 
Li$_2$ZrCuO$_4$
with a different set 
will {\it not} change this fundamental interrelation.

To compare the $J_i$
with those from another method, density
functional theory (DFT) calculations
for Li$_2$CuO$_2$ using the structural data of Ref.~5 were performed
in the local spin density plus Hubbard $U$ (LSDA+$U$)
approach. The full-potential local-orbital (FPLO)
code \cite{koepernik} was employed. The onsite Coulomb integral
for a Cu 3$d_{xy}$ hole was taken to be $U$~=~8~eV and an onsite
exchange integral $J$~=~1~eV was used. The LSDA+$U$  yields quite
reliable results for total energy differences of different spin 
structures from which the $J_i$-values of Eq.~(1) can 
be extracted. TAB.~I 
\begin{table}[]
\caption{\label{tab} In-chain exchange
integrals $J_{i}$ and frustration ratio $\alpha=-J_2/J_1$
obtained from exact diagonalizations (ED) of the Hubbard 
model (2)
with sets M2 and M1 
 as well as
from {\it independent}
 DFT+$U$ and a quantum chemistry (QC) 
study. 
}
\begin{tabular}{|l|c|c|c|c|c|}
\hline
&\multicolumn{2}{|c|}{present work}& \multicolumn{3}{|c|}{other work}\\
& ED$^a$&DFT$^b$&ED$^c$&
DFT$^d$&QC$^e$\\ \hline
$J_{1}$, K&-146 &-154& -103&-126&-142\\
$J_{2}$, K&33&30&49&52&22\\
$J_{3}$, K&-0.5&--&-2&--&--\\
$\alpha$&0.23&0.19&0.47&0.41&0.15\\
\hline
\hline
\end{tabular}\\
\begin{flushleft}
{\footnotesize 
$^a$ M2, Cu$_4$O$_{10}$ chain cluster \hfill{}\\
$^b$ LSDA+$U$ (FPLO), $U=8$ eV, $J=1$ eV\hfill{}\\
$^c$ M1 [1]; Cu$_4$O$_{10}$ chain cluster \hfill{}\\
$^d$ GGA+$U$ (WIEN2$k$); 
$U_{\mbox{\tiny eff}}=U-J=8$~eV, see Ref.~14\hfill{}\\
$^e$ TAB.~II of  
Ref.~9}\hfill{}\\
\end{flushleft}
\end{table}
shows the $J_1$- and $J_2$-values 
we found in comparison with similar 
results given in Refs.~\onlinecite{Whangbo07,Graaf02}
incl.~ a quantum chemistry study of di- and 
trimers,  
with those  from projecting   the models M1 and M2 
onto Eq.\ (1).
The small $J_3$-values justify
a description of Li$_2$CuO$_2$ in terms of a $J_1$-$J_2$ model.
There is an almost perfect
agreement of three 
independent 
approaches to be close to $\alpha_c$~=~0.25 
($J_3$~=~0) of a quasi-1D situation. 
Moreover, model M2, 
the LSDA+$U$ as well as  
the QC calculations point to
$\alpha$~$<$~$\alpha_c$, i.e.\ to a FM chain.
This holds 
in our LSDA+$U$, 
if 7~eV~$\leq U \leq$~9~eV.
\indent
To conclude, 
we have
shown that the model many-body 
Hamiltonian (2) 
with parameters M2 is capable of consistently describing XAS, EELS, RIXS 
and $\sigma(\omega)$ of Li$_2$CuO$_2$ at 300~K. The 
main issue, 
however, is that the magnetic 
\textcolor{black}{GS}
and the 
spin 
excitation spectrum strongly affect the $T$-dependence of 
$\sigma(\omega)$
in the visible range. 
The reason for is the 
thermal population of excited spin states
which differ magnetically much from the GS.
This 
qualitatively new effect, 
{\it irrespectively} on details of the
relevant microscopic parameters  for a particular compound,
has been overlooked so far in the  
cuprate optics literature. Our findings allow a quick qualitative 
magnetic classification
of cuprates 
by optical measurements:
if the AFM (FM) exchange is dominant,
$\sigma(\omega)$ 
in the ZRS energy region
increases (decreases) lowering $T$.
In the 1D FM-$J_1$--AFM-$J_2$ case 
it even allows a 
sharp determination of the position of a system 
relative to the critical point.
Thus, a $T$-dependent
$\sigma(\omega)$ 
owing to thermally activated
spin excitations is a 
general 
phenomenon. 
Experimental studies
of this $T$-dependence
together with a quantitative theoretical analysis
of
many-body models
can provide deep 
insight in the GS and the spin excitation
spectrum of undoped insulators.
This is relevant also 
for weakly coupled systems with un-shared CuO$_4$-plaquettes 
such as Bi$_2$CuO$_4$\cite{janson} or 
Sr$_2$Cu(PO)$_4$ \cite{johannes}.  
For Li$_2$CuO$_2$ 
such studies can be decisive 
to settle the intra-chain exchange and 
the 
reason for the FM in-chain order. 
If $\alpha_c >\alpha$ would 
be true 
and the FM in-chain order below $T_N$ would be due to 
weak inter-chain exchange, a {\it non-monotonic} behavior of 
$\sigma(\omega,T)$
above $T_N$ should be observed. 
For La$_6$Ca$_{8}$Cu$_{24}$O$_{41}$ and Y$_2$Ca$_2$Cu$_5$O$_{10}$  also showing 
FM in-chain order a similar situation can be anticipated. 
Near a critical point as in Li$_2$ZrCuO$_4$
 \cite{drechsler07b} 
also the 
magnetic field 
dependence becomes an issue.

We thank the DFG (Dre-269/3-1;D,M),
its Emmy-Noether program~(R), 
the ASCR project AVOZ10100520 (M),
the GIF (M,R,E) for support,  
and  M.~Knupfer, R.~Kuzian, 
J.~Richter, and S.~Nishimoto 
for discussions.

{\it Note added.} After submission of the present paper 
$\sigma(\omega,T)$  for Sr$_2$CuO$_3$
has been studied by Kim {\it et al.}
(arXiv:0806.2083). There (see Fig.~1) a slight reduction of the intensity of the 
charge transfer transition (CT) near 1.8~eV by raising $T$ has been found,
similarly to what we have shown in Fig.~3. In our approach  
this behavior is attributed to optical excitations from 
thermally 
populated 
low-lying states which exhibit a reduced $\sigma(\omega)$ in the  
mentioned above CT region. For long chains the existence of such states 
(e.g.\ triplets) {\it independently} on the value of $J_1>0$ follows 
from the Bethe-ansatz solution of the related 1D-AFM spin-1/2 Heisenberg
model.


\begin{thebibliography}{99}

\bibitem[*]{dre}Corr.\ author; electr.\ address: 
drechsler@ifw-dresden.de

\bibitem{mizuno}Y.~Mizuno, T.~Tohyama, S.~Maekawa {\it et al.}, Phys.\ Rev.\ B 
 {\bf 57}, 5326 (1998).
\bibitem{atzkern} S.~Atzkern, M.~Knupfer, M.~Golden {\it et al.}, ibid. 
{\bf 62}, 7845
(2000).
\bibitem{drechsler07a} 
S.-L.~Drechsler, J.~Richter, J.~M\'alek, 
{\it et al.}, J.\ Mag.\ \& Mag.\ Mat.\ {\bf 290}, 
345 (2005); 
J.\ of Phys.\ Cond.\ Mat. {\bf 19},  145230 (2007).

\bibitem{Vernay07}F.\ Vernay, B.~Moritz, I.~Efimov 
{\it et al.}, 
Phys.~Rev.~B {\bf 77}, 104519 (2008).
\bibitem{Learmonth07}T.~Learmonth, C.~McGuinness, P.A.~Glans {\it et al.},  
Europhys.~Lett.\  {\bf 79}, 47012 (2007).

\bibitem{sapina}F.~Sapina, J.~Rodriguezcarvajal, M.J.~Sanchis {\it et al.}, 
Sol.\ Stat.\ Comm.\ 
{\bf 74}, 779 (1990).
\bibitem{neudert}R.\ Neudert, H.~Rosner, S.-L.Drechsler {\it et al.}, 
Phys.\ Rev.\ B {\bf 60}, 13413 (1999).
\bibitem{hu}Z.\ Hu, S.-L.\ Drechsler, J.\ M\'alek 
{\it et al.}, \textcolor{black}{Europhys.\ Lett.\  {\bf 59},
 135 (2002).}
\bibitem{Graaf02}C.~de Graaf, I.~de~P.~R.~Moreira, F.~Illas, O.~Iglesias, and
A.~Labarta, 
Phys.\ Rev.\ B {\bf 66}, 014448 (2002).
\bibitem{Hasan03}M.~Hasan, Y.~Li, Y.~Chuang {\it et al.}, Int.\ J.\ of Mod.\ Phys.\ {\bf 17}, 3519
(2003).
\bibitem{Kim04}Y.~Kim, J.~Hill, F.~Chou {\it et al.}, Phys.\ Rev.\  B  {\bf 69}, 155105 (2004).
\bibitem{Chung03}E.M.L.~Chung, G.J.~McIntyre, D.M.~Paul, G.~Balakrishnan, and
M.R.~Lees, 
ibid.\ {\bf 68}, 
144410 (2003).
\bibitem{drechsler07b}S.-L.\ Drechsler {\it et al.},  
Phys.\ Rev.\ Lett.\  
{\bf 98}, 077202  (2007).


\bibitem{Whangbo07}
H.J.\ Xiang, C.\ Lee and M.-H.\ Whangbo, 
Phys.\ Rev.\ B {\bf 76}, 220411(R)  (2007).






\bibitem{jaclic}Similar finite $T$-approaches have been used for 
the $t$-$J$ model \textcolor{black}{e.g.}
by J.\ Jakli\v{c} 
and P.\ Prelov\v{s}ek, Adv.\ 
Phys.\ {\bf 19}, 1 (2000), for 
spin-less fermions  by
 X.~Zotos and P.~Prelov\v{s}ek, Phys.\ 
Rev.\ B {\bf 53}, 983 (1996), and for   
single-band  Hubbard models at $U \gg t $ 
by H.\ Onodera,  
T.~Tohyama, and S.~Maekawa, ibid.~{\bf 69}, 245117 (2004).
\bibitem{remarkenergy}This notation reflects properly only 
the excitation energies. There is no
change of $S_z$ or $S$ of a chain (cluster) 
as the term "triplet" excitation might suggest. 
\bibitem{Bondino07}F.~Bondino, M.~Zangrando, M. Zacchigna {\it et al.}, Phys.\ Rev.\ B {\bf 75},
195106 (2007).
\bibitem{Pagliara02}S.~Pagliara, F.~Parmigiani, P.~Galinetto,  
A.~Revcolevschi, and G.~Samoggia, ibid.\ {\bf 66}, 
024518 (2002).
\bibitem{koepernik}K.~Koepernik and H.~Eschrig,
ibid.~{\bf 59}, 1743 (1999). 
Here we used the code FPLO 5.00-19 and the
basis set of Ref.~\onlinecite{drechsler07b}. 

\bibitem{janson}O.~Janson, R.~Kuzian, S.-L.~Drechsler, and H.~Rosner,
ibid.~ 
{\bf 76}, 115119 (2007).
\bibitem{johannes}M.~Johannes, J.~Richter, S.-L.~Drechsler, and H.~Rosner, ibid.\
{\bf 74}, 174435 (2006). 
\end{thebibliography}
\end{document}